\documentclass[11pt, a4paper, nofootinbib]{article}
\usepackage{graphicx}
\usepackage[bookmarks, colorlinks, breaklinks, pdfauthor={Huw Price}]{hyperref}
\usepackage{tikz}
\def\R2Lurl#1#2{\mbox{\href{#1}{\tt #2}}}

\usepackage{amsmath, amsthm, amssymb}

\usepackage{sectsty}
\subsubsectionfont{\hspace{-0.60cm}
\normalfont\emph}
\subsectionfont{\hspace{-0.60cm}\nohang\normalfont\normalsize\textit}
\sectionfont{\nohang\centering\normalfont\textsc}

\newcommand{\comment}[1]{}

\hypersetup{linkcolor=blue,citecolor=blue,filecolor=black,urlcolor=blue}

\def\str#1{{\setbox1=\hbox{#1}\leavevmode 
\raise.47ex\rlap{\hspace{4.5pt}\leaders\hrule\hskip\wd 1} 
\box1}}

               \usepackage{latexsym}

\DeclareRobustCommand\textprime{\leavevmode \raise.8ex\hbox{\text@char\scriptfont\prime}}

\newcommand{\figneweprb}{\begin{center}
\begin{tikzpicture}
\begin{scope}[scale=1.9]
\draw[color=black,ultra thin] (-3.2,-1)--(-3.2,3.2)--(3.2,3.2)--(3.2,-1)--(-3.2,-1); 
\draw[color=black, ->] (2.5,0)--(2.5,1); 
\draw[color=gray,ultra thick] (2.25,1.75)--(1.75,1.25)--(1.25,1.75)--(1.75,2.25)--(2.25,1.75)--(1.75,1.25);
\draw[color=gray,thick] (1.75,1.4)--(1.75,2.1);

\draw[color=blue,densely dashed] (2,2)--(2.5,2.5);
\draw[color=blue,densely dashed] (1.5,2)--(1.0,2.5);

\draw[color=gray,ultra thick] (-2.25,1.75)--(-1.75,1.25)--(-1.25,1.75)--(-1.75,2.25)--(-2.25,1.75)--(-1.75,1.25);
\draw[color=gray,thick] (-1.75,1.4)--(-1.75,2.1);

\draw[color=blue,densely dashed] (-2,2)--(-2.5,2.5);
\draw[color=blue,densely dashed] (-1,2.5)--(-1.5,2);

\draw (-1,2.7) node {{\small $\textup{\textbf{A}}=0$}}; 
\draw (-2.5,2.7) node {{\small $\textup{\textbf{A}}=1$}};
\draw (2.5,2.7) node {{\small $\textup{\textbf{B}}=1$}};
\draw (1,2.7) node {{\small $\textup{\textbf{B}}=0$}};

\draw[color=blue,thick] (0,0)--(1.5,1.5); 
\draw[color=blue,thick] (-1.5,1.5)--(0,0); 

\draw (0,-1.5) node {{{\bf Figure 1.} EPR experiment with photons.}};
\draw (0,-0.15) node {{\small{Source}}};
\draw (0.6,0.76) node  [rotate=45] {{\small Photon}};
\draw (-0.6,0.76) node  [rotate=-45] {{\small Photon}};
\draw (2.35,0.5) node  [rotate=90] {{\small Time}};

\draw (1,1.75) node {{\small\parbox{2cm}{{Bob sets\\ angle $\beta$}}}};
\draw (-0.5,1.75) node {{\small\parbox{2cm}{Alice sets\\ angle $\alpha$}}};

\draw (0,-0.7) node {{Correlation: $Pr(\textup{\textbf{A}}=\textup{\textbf{B}}) =\cos^{2}(\alpha -\beta)$}};

\end{scope}

\end{tikzpicture}
\end{center}}

\newcommand{\newfigthree}{\begin{center}
\begin{tikzpicture}
\begin{scope}[scale=1.9]

\draw[red,fill=orange] (1.2,1.2) circle (.4ex); 
\draw[red,fill=orange] (-1.2,-1.2) circle (.4ex); 

\draw[color=black,ultra thin] (-3.2,-3.6)--(-3.2,3.2)--(3.2,3.2)--(3.2,-3.6)--(-3.2,-3.6); 
\draw[color=black, ->] (2.5,-2)--(2.5,1); 
\draw[color=gray,ultra thick] (2.25,1.75)--(1.75,1.25)--(1.25,1.75)--(1.75,2.25)--(2.25,1.75)--(1.75,1.25);
\draw[color=gray,thick] (1.75,1.4)--(1.75,2.1);

\draw[color=blue,densely dashed] (2,2)--(2.5,2.5);
\draw[color=blue,densely dashed] (1.5,2)--(1.0,2.5);

\draw[color=gray,ultra thick] (-2.25,-1.75)--(-1.75,-1.25)--(-1.25,-1.75)--(-1.75,-2.25)--(-2.25,-1.75)--(-1.75,-1.25);
\draw[color=gray,thick] (-1.75,-1.4)--(-1.75,-2.1);
\draw[color=blue,densely dashed] (-2,-2)--(-2.5,-2.5);
\draw[color=blue,densely dashed] (-1,-2.5)--(-1.5,-2);
\draw (-1,-2.7) node {{\small $\textup{\textbf{A\ensuremath{'}}}=0$}}; 
\draw (-2.5,-2.7) node {{\small $\textup{\textbf{A\ensuremath{'}}}=1$}};
\draw (2.5,2.7) node {{\small $\textup{\textbf{B}}=1$}};
\draw (1,2.7) node {{\small $\textup{\textbf{B}}=0$}};

\draw[color=blue,thick] (0,0)--(1.5,1.5); 
\draw[color=blue,thick] (-1.5,-1.5)--(0,0); 

\draw (0.,-4.1) node {{\bf Figure 3.} The one-photon experiment.};
\draw (0.6,0.76) node  [rotate=45] {{\small Photon}};
\draw (2.35,-0.5) node  [rotate=90] {{\small Time}};

\draw (1,1.75) node {{\small\parbox{2cm}{{Bob sets\\ angle $\beta$}}}};
\draw (-0.5,-1.75) node {{\small\parbox{2cm}{Ecila sets\\ angle $\alpha$}}};
\draw (-1.77,-3.1) node {{\small Erutan chooses an input}};
\draw (1.75,3) node {{\small Nature chooses an output}};
\draw (-0.1,-1.0) node {{\small\parbox{2cm}{$\tau$}}};

\end{scope}

\end{tikzpicture}
\end{center}}
    
\begin{document}  

\title{Disentangling the Quantum World}
\author{Huw Price\thanks{Trinity College, Cambridge CB2 1TQ, UK; email \href{mailto:hp331@cam.ac.uk}{hp331@cam.ac.uk}.} {\ and} Ken Wharton\thanks{Department of Physics and Astronomy, San Jos\'{e} State University, San Jos\'{e}, CA 95192-0106, USA; email \href{mailto:kenneth.wharton@sjsu.edu}{kenneth.wharton@sjsu.edu}.}}
\date{}  
\maketitle
\begin{abstract}
\noindent Correlations related to quantum entanglement have convinced many physicists that there must be some at-a-distance connection between separated events, at the quantum level. In the late 1940s, however, O.~Costa de Beauregard proposed that such correlations can be explained without action at a distance, so long as the influence takes a zigzag path, via the intersecting past lightcones of the events in question. Costa de Beauregard's proposal is related to what has come to be called the retrocausal loophole in Bell's Theorem, but -- like that loophole -- it receives little attention, and remains poorly understood. Here we propose a new way to explain and motivate the idea. We exploit some simple symmetries to show how Costa de Beauregard's zigzag needs to work, to explain the correlations at the core of Bell's Theorem. As a bonus, the explanation shows how entanglement might be a much simpler matter than the orthodox view assumes -- not a puzzling feature of quantum reality itself, but an entirely unpuzzling feature of our knowledge of reality, once  zigzags are in play.   
\end{abstract}









\section{Strange connections}

One of the most puzzling things about quantum mechanics (QM) is \emph{entanglement} -- the strange connection between quantum systems that  allows each to know something about what's happening to the other, no matter how far apart they may be. Erwin Schr\"odinger, who invented the term, said that entanglement was not just one but ``rather \emph{the} characteristic trait of quantum mechanics, the one that enforces its entire departure from classical lines of thought.''\cite{ES}

Schr\"odinger was discussing so-called EPR experiments, invented by Einstein, Podolsky and Rosen (EPR) in 1935. A typical case is shown in Figure 1. Two particles (photons, in this version) are created together at a source, and sent in different directions to experimenters Alice and Bob, who each have a choice of several measurements they can perform on their particle. And although each outcome is unpredictable on its own, when they choose matching measurements, the particles turn out to be perfectly correlated: the two outcomes match 100\% of the time. EPR used these correlations to argue that the particles must carry ``hidden instructions'', telling the particles how to behave for each measurement that Alice and Bob might choose to perform.  They concluded that standard QM was incomplete, because it didn't describe these hidden instructions.\cite{EPR}

\begin{figure}[t]
\figneweprb
\end{figure}

If EPR had been right about hidden instructions, quantum correlations would be no more spooky than similarities between identical twins who share the same genetic ``instructions". But in 1964 John Bell proved that the quantum case is different. Bell's Theorem shows that any hidden instructions would themselves have to rely on action at a distance, to be consistent with the predictions of QM. (Many experiments have since confirmed these predictions.)  

Entanglement is this counterintuitive connection between Alice's particle and Bob's, somehow able to guarantee certain correlations, no matter how far apart Alice and Bob might be. They could be separated by lightyears, or have the mass of a planet between them, but entanglement doesn't care. Whether this is explained by a framework in which separated objects aren't truly separated,\footnote{E.g.,  as in orthodox QM,  a framework in which particles no longer exist in ordinary space, but instead are linked directly to other particles in some vastly higher-dimensional structure.}  or by allowing instantaneous communication at the level of hidden parameters, this distant connection is termed ``nonlocality''. And it is now widely believed to be essential to the quantum world. 

Nonlocality imperils more than just our sensibilities about action at a distance.  As David Albert and 
Rivka Gal\-chen put it,  in a recent
piece in \emph{Scientific American}: ``Quantum mechanics has upended many an intuition,
but none deeper than [locality]. And this particular upending carries
with it a threat, as yet unresolved, to special relativity---a
foundation stone of our 21st-century physics.''\cite{Albert}

\section{The Parisian Zigzag}

Back in 1935, thirty years before Bell's Theorem, it still seemed ``obvious'' that there could be no action at a distance. As Schr\"odinger put it at that time, ``measurements on separated systems cannot directly influence each other -- that would be magic.''\cite{ES2} The EPR argument for ``hidden instructions'' assumed that Alice's choice of measurement cannot influence Bob's particle, and vice versa. 

But a  decade later, in post-war Paris, a young French graduate student, Olivier Costa de Beauregard, spotted an interesting loophole in EPR's reasoning. He realized that Alice's choices could affect Bob's particle \emph{indirectly} -- {so without action at a distance} -- if the effect followed a zigzag path, \emph{via the past.} Alice's choice could affect her particle ``retrocausally'', so to speak, right back to the common source, in turn correlating Bob's particle with Alice's choice (and vice versa).\cite{OCB}

Unfortunately for Costa de Beauregard, his thesis advisor was Louis de Broglie, one of the giants of early quantum theory (and a prince, to boot!) For several years, de Broglie forbade his student to publish his strange idea -- relenting only when Feynman published a famous paper describing positrons as electrons zigzagging backwards in time.  Despite the Feynman factor, however, Costa de Beauregard's proposal made no impact among the Copenhagen-minded physicists of the day. (Most of them thought that Bohr had already dealt with the EPR argument.)

Ironically, one of the few anti-Copenhagen heretics in those days was the young John Bell, whose conviction that EPR were making an important point was to lead him to his own famous reason for thinking that Einstein must nevertheless be wrong. As Bell himself put it, many years later: ``For me it is so reasonable to assume that the photons in those experiments carry with them programs [i.e., ``hidden instructions''], which have been correlated in advance,  telling them how to behave. This is so rational that I think that when Einstein saw that, and the others refused to see it, he was the rational man. The other people, although history has justified them,  were burying their heads in the sand. \ldots\  So for me, it is a pity that Einstein's idea doesn't work. The reasonable thing just doesn't work.''\cite{BellQuote}

In the decades after Bell's Theorem, a few writers noticed that Costa de Beauregard's loophole also applied to Bell's reasoning.  As Bell himself makes clear, his result requires the assumption that Alice and Bob's measurement choices \emph{don't} affect the prior state of the particles. If we allow such retrocausality -- if Alice's and Bob's choices affect their particles' common past -- then Bell's argument for action at a distance is undermined.

This loophole receives  little attention, and remains  poorly understood. Our purpose here is to throw some light on the idea, by proposing a new way to explain and motivate it. We exploit some simple symmetries to show how Costa de Beauregard's zigzag would need to work, to explain the correlations at the core of Bell's Theorem. As a bonus, the explanation shows how entanglement might be a much simpler matter than the orthodox view assumes -- less a puzzling feature of quantum reality itself, than an entirely unpuzzling feature of our \emph{knowledge} of reality, once the zigzags are in play. It also shows how one of the intuitive objections to retrocausality -- that it would lead to time-travel-like paradoxes -- can be avoided very easily. 

An important note about terminology, before we begin. The central idea of  Costa de Beauregard's proposal is that Alice's choices may affect what happens on Bob's side of the experiment, without action at a distance, so long as the effect goes via the past. Does this mean that the zigzag avoids {nonlocality?} Yes in one sense, but no in another, \emph{for the term `locality' is {ambiguous,} once the zigzag option is in play.} If `local' means that Alice's choices can't affect Bob's simultaneous measurement \emph{at all,} then the zigzag model is \emph{not} local. But if it means that every distant influence must be explained by some contiguous chain of intermediate events in space-time (with no faster-than-light individual links), then zigzags \emph{are} local. 

Normally, these two meanings of `locality' would be thought to coincide, but they come apart in zigzag models. To avoid confusion, we shall avoid the term altogether, from now on. But we note that it is the second sense of nonlocality -- fundamental faster-than-light processes -- that offends both old objections to action at a distance and new objections based on relativity. A great advantage of the Parisian zigzag, if it works, is that it avoids nonlocality in this sense.

\section{Alice through the looking glass}

\subsection{Polarizing cubes}

Let's begin with some more details about the kind of experiment depicted in Figure 1. The devices that Alice and Bob control are intended to be polarizing cubes.  Classically, such cubes separate the polarization components of the incoming light.  Each cube can be set at an arbitrary angle, and any incoming light whose polarization matches the chosen angle will pass straight through.  But any incoming light with a polarization perpendicular to this chosen angle will reflect off the line drawn in the center of the cube, and change direction. 

Generally, then, in the classical case, the cube will split one incoming beam into two outgoing beams. The exceptions are the cases in which incoming beam is already polarized along (or perpendicular to) the setting angle of the cube. In those cases, 100\% of the outgoing light lies in a single beam. 

For future reference, we note that  such a cube can also be used in reverse, to ``splice'' two suitably polarized (and phase-locked) beams into a single beam. Splicing is simply the time-reverse of splitting, and its possibility follows from the fact that  classical electromagnetism is time-symmetric.

However, this ``splitting''  of classical electromagnetic waves does not extend to the low-energy limit introduced in quantum theory. In this limit,  we  encounter particle-like packets called ``photons''. When the experiment of Figure 1 is conducted with a single pair of photons, each of these photons is found entirely on one path or the other, at the relevant  wing of the experiment.\footnote{The  \emph{probability} of finding a photon on each path may still be said to ``split'', thus matching classical predictions in the many-photon limit, but our concern will be in making sense of these experiments at the level of single photons.} This difference between the quantum and classical cases is crucial to entanglement and the case for nonlocality. Bell's Theorem, for example, turns entirely on the correlations between these ``discrete'' single-photon outcomes, on the two sides of experiments such as that of Figure 1. There are no such discrete phenomena in the classical case.

For fully entangled photons, as in Figure 1, the strange correlations between Alice's and Bob's outcomes are masked by a curious feature: each individual outcome appears to be completely random.  No matter what setting Alice chooses, she always finds a 50\% chance of measuring her photon on each of her two possible outputs ($A$). The same goes for Bob's outcome $B$.  It's only when they compare notes, after the fact, that the strange correlations become apparent.

\subsection{Into the mirror}

With these preliminaries in place, let's now stand Figure 1 on top of a mirror, as shown in Figure 2. 
In the mirror we can see the reflection of Alice and her half of the experiment. (There is no mirror under Bob's half.)  Now focus on this reflection of Alice and her half of the experiment, and combine it in your mind's eye with Bob's half of the original figure. This combination (reflected Alice, plus Bob) looks exactly like the  spacetime diagram of a different experiment -- one  in which a single photon passes from Alice to Bob, going through two polarizing cubes.\footnote{To make this trick work, we have to be careful to  place the far righthand corner of the mirror at the point on Figure 1 where the entangled particles are created.}  

\begin{figure}[t]
\begin{center}
\includegraphics[height=11.2cm]{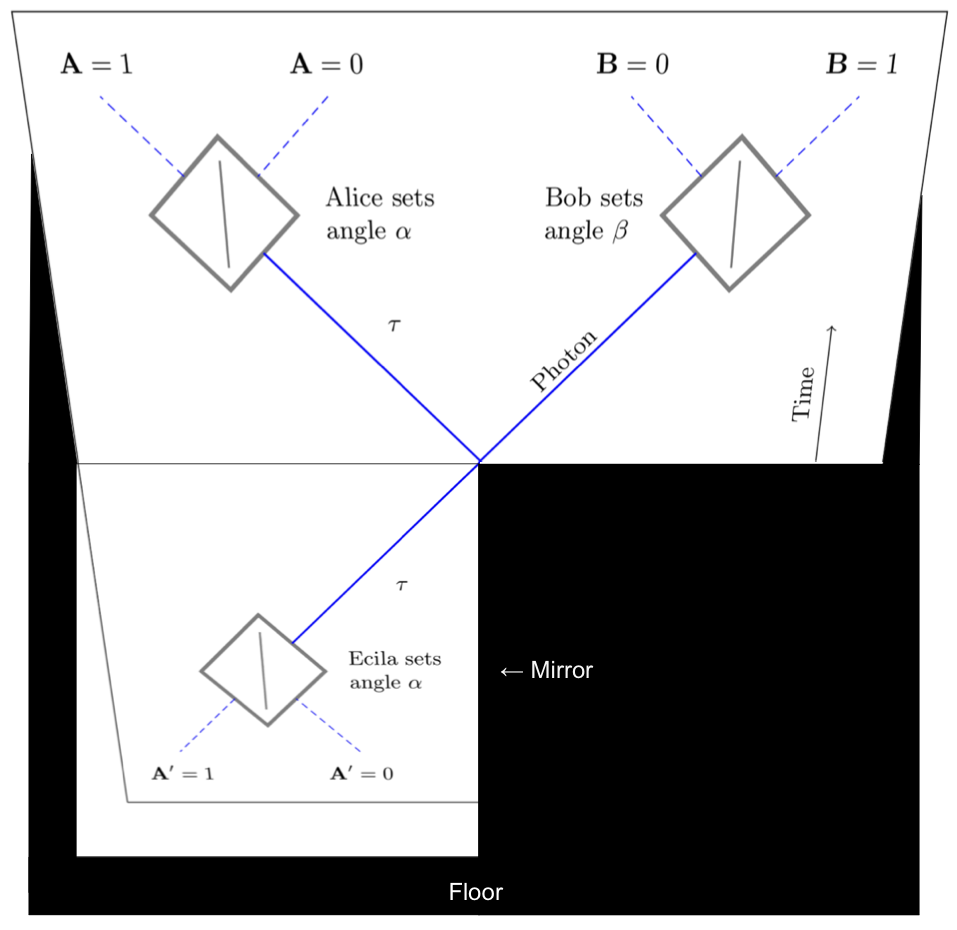}\\
{\bf Figure 2.} Alice through the looking glass.
\end{center}
\end{figure}

To avoid confusion we've reproduced this new one-photon experiment as Figure 3. Reflection in the mirror corresponds to time-reversal, so we have named Alice's time-reversed counterpart `Ecila'. (Ignore the orange dots for the moment.)

\subsection{The one-photon experiment}

This new experiment (Figure 3) is not some impossible permutation of the original -- it's a perfectly valid experiment, which we can actually carry out.  But it is unusual in one respect, and this oddity will play a crucial role in the use we want to make of the experiment. Normally, if we were performing a two-polarizer experiment of this kind, it would be natural to take advantage of our ability to control the input channel ($A'$) at Ecila's end of the experiment. We (or Ecila herself) could simply choose to supply photons on one channel or the other.

For Ecila's end of Figure 3 be a proper mirror image of Alice's end of Figure 1, we need to do something different. We need to ensure that photons are secretly supplied \emph{at random} on one channel or other, to mirror the unpredictable random outputs from Alice's cube in Figure 1. We call this random source `Erutan', since it mirrors the action of Nature in choosing Alice's outputs in Figure 1.\footnote{In Figure 1, we are stipulating that Alice is measuring which output path ($A$) the photon ends up on. To mirror this behaviour in Figure 3, it must also be a determinate matter that Ecila's input photon arrives on one input path ($A'$) or other.}\\  

\newfigthree

Intriguingly, the correlations that one sees between Ecila's inputs and Bob's outputs in real-life versions of Figure 3 are exactly the same as those between Alice's outputs and Bob's outputs in Figure 1 -- and they depend on the choice of $\alpha$ and $\beta$ in exactly the same way.
In the case of Figure 3, however, the explanation of these correlations is thought to be straightforward. The photon polarization $\tau$ is determined both by Erutan's randomly-chosen input at Ecila's end of the experiment and by Ecila's choice of the measurement setting $\alpha$.  (More on the details of this below.) In turn, $\tau$ makes a difference to the outcome at Bob's end of the experiment, in combination with Bob's setting $\beta$. 

Intuitively, the polarization $\tau$ connects events at one end to events at the other, without action at a (temporal) distance, or any mysterious entanglement. We simply have a single enduring property of the photon, $\tau$, that `bridges the temporal gap', and ensures in an entirely non-mysterious fashion that the output channel of the photon at Bob's end of the experiment is related to its input at Ecila's end of the experiment, in a way that depends on the settings chosen by Ecila and Bob.

\subsection{Backing out of the mirror}

We want to use this simple, uncontroversial explanation from Figure 3 to put some flesh on the bones of Costa de Beauregard's zigzag, in Figure 1.  More precisely, we want to reach into the mirror in Figure 2, pull out the $\tau$-based explanation, and apply it to the original EPR experiment in Figure 1.  We are exploiting the fact that, in effect, the mirror is already showing us \emph{exactly} what we need for a zigzag explanation of the correlations in Figure 1. We simply need to assign a property $\tau$ to the photon in Figure 1, \emph{before} it reaches Alice's cube (matching the property of the photon in Figure 3, \emph{after} it leaves Ecila's cube).  

If we allow this property $\tau$ to be a constant throughout the zigzag path from Alice to Bob in Figure 1 (just as it is a constant from Ecila to Bob in Figure 3), then it plays exactly the same role in ``propagating influence'' to Bob in one experiment as in the other. That is, it makes exactly the same contribution to showing how a difference in Alice's choice of the setting $\alpha$ makes a difference to the photon in the region of Bob's cube, as it does to showing how a difference in \emph{Ecila's} choice of the setting $\alpha$ makes a difference to the photon in the region of Bob's cube. (For philosophers we might say: The relevant counterfactuals are exactly the same!)  So we have an explanation -- or, for cautious folk, an ``explanation''! -- of the correlations in Figure 1, just as we do in Figure 3. 

So we now have a picture of an EPR-style experiment that shows us how the world needs  to behave, to explain the Alice--Bob correlations via Costa de Beauregard's ``retrocausal'' proposal. Want to know what the retrocausality needs to look like? Just think about what ordinary causality looks like in Figure 3, according to the standard quantum picture, in the special case in which the input channel is random. The control that the retrocausal proposal needs to give to Alice is exactly the control that looks like the standard ``forward causal'' story, when reflected in the mirror. We want to see  Ecila controlling $\tau$ after it leaves her cube (that's the standard story),  so we need to show Alice controlling  $\tau$ \emph{before} it reaches her cube -- and that's the retrocausality, in the zigzag proposal.  

\subsection{Too good to be true?}

At this point, readers may feel that our use of the mirror involves some sort of sleight of hand, and that there are obvious difficulties for the zigzag proposal. We can't anticipate all such concerns,  but we want to respond to three objections: \begin{enumerate}
\item If Alice can control $\tau$ over on Bob's side of the experiment, why can't she send a signal to him? It is well-known that QM does not allow signalling  in the kind of experiment depicted in Figure 1,  and we might therefore suspect that this zigzag connection would be incompatible with standard QM. (Typical causal channels can be used to signal, after all.) 
\item Why is Alice allowed to influence $\tau$, when Bob seems to do no such thing? (Discrimination against experimenters on the right!)
\item Isn't the zigzag model just another version of the discredited ``superdeterminist'' proposal?
\end{enumerate}

\noindent We'll come back to (2)  in due course, and show how the zigzag proposal can give Bob an influence, too; and we'll come back to (3), say what it means, and show why the objection is mistaken.  But first let's explain why Alice can't signal to Bob, even if she has retrocausal control of $\tau$. To do this, we need to go back into the mirror.

\section{Causation without signalling}

\subsection{Can Ecila signal?}

We are interested in whether Alice could signal to Bob in Figure 1, if she had retrocausal control of $\tau$. To answer this question, let us first ask the corresponding question about Ecila. Is it possible for Ecila  to signal to Bob in Figure 3? 

If Ecila could control the path chosen by the input photon, the answer would be certainly be `Yes'. Fixing the path for a series of runs of the experiment would give Ecila complete control over the polarization $\tau$, and Ecila could then vary $\tau$ to encode a message -- she simply needs to send enough photons with the same polarization, for each bit of her message, to enable Bob to measure the polarization. 

Suppose, for example, that Ecila's photons come reliably from the lower left channel. To ensure that some photons sent to Bob have polarization $\tau$, Ecila sets $\alpha = \tau$. Bob experiments with various settings $\beta$, and discovers that the bias between his outputs is greatest when $\beta = \alpha$ or $\beta = \alpha+90^\circ$, and disappears altogether when $\beta = \alpha+45^\circ$. This tells him that  $\tau = \alpha$ or $\tau = \alpha+90^\circ$ (and the direction of the bias between outputs will distinguish these possibilities). 

Signalling remains possible even if Ecila doesn't know which channel her photons are arriving from, so long as they all arriving from the same channel. In fact, it is enough that there is a reliable bias, so that one channel reliably has higher probability than the other. Bob can still detect that Ecila's setting is either 
$\alpha$ or $\alpha+90^\circ$, by looking for the setting of his polarizer that produces maximum bias in his outputs. And this information is enough to carry a signal. (For example, ``$\alpha = 0^\circ$ or $ \alpha = 90^\circ$'' could mean ``0'', while ``$\alpha = 45^\circ$ or $ \alpha = 135^\circ$'' means ``1''.)

However, when Ecila's input photon is supplied by a hidden randomizer, with no bias between the two input channels, this kind of signalling becomes impossible.  (As we have specified above, the randomizer Erutan acts as a mirror image of Nature.)   If Ecila chooses setting $\alpha$ in this case, each photon sent to Bob has equal probability of having polarization $\alpha$ (if the input came from the lower left), or $\alpha + 90^\circ$ (if the input came from the lower right).  Even if Bob happens to set $\beta = \alpha$ he cannot tell that he has done so, because his outputs display no bias, thanks to the random and hidden input at Ecila's end of the experiment.

With the randomizer in place, then, Ecila's choice of the angle $\alpha$ does not give her enough control to send a signal to Bob. It might be thought that she has lost control of the polarization $\tau$ altogether, but this is not so. She retains enough control to restrict the photon to just two possible polarizations ($\alpha$ or $\alpha + 90^\circ$). Intuitively, then, Ecila has a \emph{causal influence} on the polarization $\tau$, without being able to use that influence to signal to the future -- and the fact that she still has causal influence continues to play a crucial role in the intuitive explanation of the correlations that obtain between her inputs and Bob's outputs.

\subsection{No signalling, with mirrors}

Now that we know why Ecila can't signal to Bob, despite having some control of $\tau$, we can see why the same is true of Alice. The control of $\tau$ that the zigzag model gives to Alice is \emph{exactly} the control of $\tau$ that Ecila retains in Figure 3.  As in that case, it is control, but it doesn't permit signalling.  So there is no conflict on this score between the zigzag model and the prohibition of signalling in Figure 1 in orthodox QM.\footnote{Note that since Ecila's inability to signal is linked to the fact that Erutan supplies symmetric inputs, an analogous claim must be true of Alice. In the particular two-photon experiment to which the mirror symmetry applies, the prohibition of signalling in orthodox QM must be linked to the fact that the Born Rule guarantees that Alice's outputs are similarly symmetric.}   (This explicit causation-without-signalling is comparable to orthodox QM's description of ``quantum steering'' in these same cases.)

\subsection{Generalised no-signalling?}

This explanation shows why Alice cannot signal in this particular two-photon experiment, despite the retrocausal control provided by the zigzag model. Can the explanation be extended to other possible entangled states?  If the zigzag model itself is to be extended to other entanglement experiments, we should hope that the no-signalling condition will also generalize.  The general issue has recently been raised by Wood and Spekkens \cite{WS}, who note that causation-without-signalling is logically possible in such causal models, but typically requires what they term ``fine-tuning''.  

In the above example, as we noted, this so-called fine tuning turns out to be naturally enforced by a symmetry between the outputs (in the case of Alice) or the inputs (in the case of Ecila).  The same symmetry has been shown to explain no-signalling for every maximally entangled state \cite{Almada}, not merely the one considered here.  An interesting open question is whether other natural symmetries (time-symmetry, Lorentz-symmetry, etc.) are available to supply the necessary ``fine-tuning'' for partially entangled states.  Resolution of this question must await a hidden variable model rich enough to encode all such states, although one promising framework can be found in \cite{WhartonKoch}.\footnote{For further discussion about how retrocausal models interface with the Wood-Spekkens fine-tuning argument, see \cite{Almada,Evans}.}

\section{What about Bob?}

As it stands, the zigzag explanation of the Alice--Bob correlations in Figure 1 shows an absurd spatial asymmetry: Alice has retrocausal control, but Bob does not. If the proposal is to have any claim to be taken seriously, this asymmetry will have to go. 

In principle, there are three ways this might be done.  One approach would be to double up the properties of the photon, so that there is a property $\tau_{a}$ controlled by Alice from the left, and  a different property $\tau_{b}$ controlled by Bob from the right. In effect, this is the approach taken by the Two State Vector approach to QM, defended by Aharonov, Vaidman and others.\cite{TSV} This can explain the Alice--Bob correlations -- it provides two consistent explanations, in fact, depending on which end we start.\footnote{Another version of this approach proposes an analogy with the Wheeler-Feynman absorber theory of radiation, thinking of $\tau_{b}$ by comparison with the advanced potential in the Wheeler-Feynman picture. See \cite{Cramer, Pegg1, Pegg2} for details.}

A second approach would be to make do with a single $\tau$ that need not be constant.  If $\tau$ is controlled on the left by Alice and on the right by Bob, it must be allowed to change value in between, from $\tau_a$ to $\tau_b$, to avoid the inconsistencies that would otherwise arise when Alice and Bob choose incompatible settings (i.e., when any single, fixed $\tau$ is incompatible with either $\alpha$ or $\beta$).  Remarkably, there is a simple rule that recovers the correct correlations between Alice and Bob for models of this sort.\cite{Wharton}

Finally, one might take the approach that the polarization $\tau$ is not a ``real'' parameter, but just a summary of the knowledge we have about the system.  This would correspond to what is now often called an `epistemic' interpretation of $\tau$.  Such an approach is not directly relevant to our present concerns.\footnote{Namely, showing how Costa de Beauregard's zigzag supports an epistemic understanding of entanglement, even if $\tau$ itself is not interpreted in an epistemic fashion.}
 However, we note that there are at least two reasons for thinking that any plausible epistemic interpretation of $\tau$ is also likely to be retrocausal, at the level of its underlying ontology. One is the requirement of time-symmetry; on this topic Pusey \cite{Pusey} has extended an argument of Price \cite{Price} to the epistemic case.  The second is that retrocausality provides one of the few loopholes in  the strongest argument against the epistemic view.\cite{PBR}
 
Setting the epistemic interpretation of $\tau$ to one side, we note that a bonus of either of the two previous approaches is that they remove a puzzling time-asymmetry in Figure 3.  This is not the asymmetry-of-signalling that was removed by Erutan, as discussed in the previous section; with Erutan present, neither Ecila nor Bob can signal to each other.\footnote{Without Erutan, Ecila knows the input channel $A'$ before she chooses her setting $\alpha$, but Bob has no such access to $B$ before he chooses his setting $\beta$.}  But even with Erutan, a further time-asymmetry remains if there's only one fixed value of $\tau$.  The standard assumption in this case is that Ecila still has control over $\tau$ in this case (up to an additive factor of $90^\circ$), while Bob does not.

For either of the time-symmetric approaches described in this section, this puzzling asymmetry disappears.  In these cases, whatever control Ecila has over $\tau_a$, Bob has the same control over $\tau_b$.  In other words, by restoring the spatial symmetry of the zigzag in Figure 1, we automatically restore the temporal symmetry of Figure 3.  This reveals that the single-fixed-$\tau$ model involves a new and apparently fundamental time-asymmetry, and it is an advantage of the zigzag models that they remove it.\footnote{This issue is new to the quantum case, being a product of the discretization in the single-photon limit. In the classical case, an initial randomizer mirroring Nature will deprive Ecila of any determinate control over the output polarization of a classical light beam at her end of such an experiment, restoring the symmetry between Ecila and Bob. For further discussion of this point, see \cite{Price}.}

\section{Isn't this just ``superdeterminism''?}

Some readers may feel that the proposal offered here is merely another version of a familiar but unpopular proposal known as ``superdeterminism''.\footnote{We are grateful to a referee at this point, and borrow her/his formulation of this challenge in this paragraph and the next.} What is superdeterminism? We can explain it by contrast to our own proposal, referring to Figure 1. On the retrocausal view, Alice's choice of angle  $\alpha$ influences a photon property $\tau$ on the segment of the photon's worldline between Alice's measurement and the point in the past where the entangled photon pair has been created.\footnote{The referee expresses this idea as the claim that the ``photon property $\tau$ \ldots\ somehow travels back in time (whatever that means).'' We agree that the notion of travelling back (or forwards) in time is problematic, and therefore avoid it. (That's why, when we spoke of influence ``propagating'' in \S3.4, we put the word in scare-quotes.) Instead we speak of properties of world-lines during temporal intervals. What is distinctive about retrocausal models, in this framework, is that they allow such properties to be influenced by choices made by experimenters at both ends of the interval.} 

The superdeterminist proposes that this can be interpreted in a different way. At the time when the photon pair was created, the random variable $\tau$ was determined, and then sent to Alice (and Bob). This element of reality \emph{forced} Alice to choose a specific setting $\alpha$ (or, in more detail, forced everything that happened close to Alice to result in a specific combination of the outcome $A$ and $\alpha$), so that Alice was not really free to choose her setting. In other words, the choice of setting was determined in the past, in such a way to conspire to make everything look like a Bell violation but satisfy no-signalling.

Superdeterminism is usually rejected for two  reasons. First, it is  felt to require either some highly implausible conspiracy in the initial conditions of the Universe, or some new realm of hidden variables with remarkable powers to control the behaviour not only of human experimenters but also of the various mechanical substitutes that might be used for choosing measurement settings, apparently at random (e.g., the Swiss national lottery machine, as Bell once proposed). Second, it is felt to conflict with ``core assumptions necessary to undertake scientific experiments.''\cite{Wiseman} As Maudlin says, ``all scientific interpretations of our observations presuppose that [our choices of settings] have not have been manipulated in such a way.''\cite{Maudlin} This objection may also be traced to Bell, who says this, for example:
\begin{quote}
``A respectable class of theories, including contemporary quantum theory as it is practised, have `free' `external' variables in addition to those internal to and conditioned by the theory. These variables \ldots\  provide a point of leverage for `free willed experimenters', if reference to such hypothetical metaphysical entities is permitted. I am inclined to pay particular attention to theories of this kind, which seem to me most simply related to our everyday way of looking at the world.''\cite{Bell}
\end{quote}
We agree with both of these objections to superdeterminism. By explaining how the retrocausal proposal avoids them, we can explain how it differs from superdeterminism. 

Taking the second objection first, the retrocausal proposal accepts the standard presupposition of all experimental science, namely that experimenters such as Alice and Bob are free to choose measurement settings. Moreover, it accepts a standard operational definition of causality -- a definition long assumed in science, and refined and formalised in philosophy over the past three decades -- in which the notion of free agency plays a central role. According to this so-called ``interventionist'' account of causality, a variable X is a cause of a variable Y if and only if a free intervention on X makes a difference to Y.\footnote{See \cite{Woodward, Pearl, MP, BJPS91} for further details, and \cite{PriceWeslake,TAAP,toy} for discussions of the application of this approach to the direction of causality and the possibility of retrocausality. A point stressed in this literature (see particularly, e.g., \cite{MP,Pearl}) is that an adequate operational definition of causation cannot be purely \emph{observational,} if observation is understood in a passive sense. In science, as in ordinary life, the notion of causation depends on the fact that  we \emph{intervene} as well as \emph{observe.}} (This approach to causality thus accords a central and indispensable role to what Bell referred to in the passage above as ``free external variables'', and vindicates his view that this notion is central to ``our everyday way of looking at the world'', as well as in science.)

From this standard interventionist definition of causation, utilizing the assumption of free control of measurement settings that superdeterminism is rightly criticized for neglecting, it follows immediately that the direction of causation in our models is the one claimed by the retrocausal reading. Alice and Bob choose the measurement settings in the normal way, and these settings in turn make a difference to the prior values of $\tau$.\footnote{Some writers object that retrocausation cannot be ``real'' causation, claiming that it is a matter of definition, or perhaps presupposed by relativistic spacetime, that true causation only works past-to-future. But this objection need not detain us here. What has happened is that two criteria for causation -- the interventionist criterion, and the time-direction criterion -- have turned out to conflict, in the kind of models here proposed. It is then a terminological choice which criterion we take to be the more important. The possibility of such a conflict  was recognised long ago by the philosopher Michael Dummett.\cite{Dummett} Dummett himself opts for the second criterion, and proposes ``quasi-causation'' for the case involving control of something in the past. In Dummett's terminology, these models for QM involve retro-\emph{quasi-causation,}  not retro\emph{causation} -- but they retain all the advantages here described, under this new name. See \cite{TAAP,PriceWeslake} for further discussion of this point.}

The retrocausal proposal also differs from superdeterminism on the points at issue in the first objection. The retrocausal model introduces no strange new hidden variables to control measurement settings. To the extent that it proposes new hidden variables, they are internal to the model (and, at least in the version referred to in these paragraphs, of a familiar kind -- e.g., a new $\tau$, controlled from the future). It may induce correlations in the initial conditions of the Universe (or subsystems of it), but if so, these will not be conspiratorial -- on the contrary, they will be explicable within the model in just the way that a standard forward-causal model explains correlations in the future.\footnote{We discuss the distinction between superdeterminism and retrocausality further in \cite{PW}.}

\section{Entanglement without spooks}

Finally, back to what we promised at the beginning: an explanation of how the Parisian zigzag offers a less spooky explanation of entanglement.  Once again, we will start with Figure 3, and then use the mirror to apply the lessons of that case to Figure 1. Please pay attention to the orange dots in Figure 3, that we earlier asked you to ignore. But now imagine that -- concerning a particular photon -- you know that it is participating in the experiment in Figure 3, but that you don't know the setting or the input/output channel, at either end. 

Consider the photon at the upper orange dot, for example, and the various possibilities for what Bob's setting and outcome might be, immediately in its future. Imagine that we are interested in the probability we should assign to each outcome, conditional on the various possible settings. What we know at this point is something rather bland: whatever the setting, the probability of each outcome is 50\%. 

This is a `subjective' or  `evidential'  probability. If we had more evidence -- in particular, if we knew the setting and input channel at Ecila's end of the experiment -- we would in general  assign a different probability to each of Bob's outcomes, for each choice of his setting.  For example, if we learn that Ecila's input and setting are $A'=1$ and $\alpha$, respectively, then we should now say that the probability of outcome $B=1$, assuming Bob chooses setting $\alpha$, is 100\%. Nevertheless, the bland 50\% probabilities are the correct probabilities, for the knowledge state we assumed here: ignorance of the setting and input, at Ecila's end of the experiment.

Exactly the same is true in reverse at the lower orange dot. There, too, the probability of each input, for each assumption about Ecila's settings, is 50\%, {if we don't know Bob's setting and outcome.}  And there, too, the probability changes, if we get additional evidence -- if we learn about the setting and outcome at Bob's end of the experiment.  

So in Figure 3 we have a perfectly time-symmetric story about how getting information about one end of the experiment affects what probabilities we are correct to assign at the other end of the experiment, in the kind of knowledge-state we assumed. These evidence-based probabilities are time-symmetric in this way, even if we are assuming that the underlying reality is not symmetric -- even if we think, as in the standard picture, that there is a real property $\tau$  influenced by Ecila but not by Bob. 

Using the mirror, we can now transfer this understanding to Figure 1. In this case, the situation in which we are ignorant of the setting and outcome at both ends looks perfectly normal, for an evident reason: they all lie in the future!  But the above analysis goes through in the very same way.  Our best description of each side of the experiment predicts bland 50\% probabilities, for each assumption about the choice of setting, until we learn about what happens {on the other side.}  At that point, we have new evidence, and  can update our probabilities on the opposite site.  

The significance of this account is that these probabilities correspond \emph{exactly} to the information carried by entanglement.  In the standard view, this ``entangled state'' is thought to be a real property, that depends in a mysterious way on what happens on the other side of the experiment -- it changes, when a measurement is made on the opposite side.  But the mirror shows us that this interpretation is not compulsory. We can understand these probabilities in terms of changing evidence, just as we did in Figure 3. 

Why is this understanding of entanglement so much harder to see in Figure 1 than in Figure 3? Because in Figure 3 we think we understand \emph{why } these evidential probabilities work the way they do -- the standard model, treating  $\tau$ as a  real property, offers  an explanation of the correlations on which these probabilities are based. In Figure 1, there doesn't seem to be any explanation on offer, except the one that thinks of entanglement in terms of a real property, mysteriously affected by choices made elsewhere. But once the zigzag model is on the table (even in the left-to-right, unfair-to-Bob version), it does the explanatory job. So it frees us to think of  entanglement in this easy, state-of-information fashion, just as we do in Figure 3. 

The project of trying to make sense of entanglement started with EPR, 80 years ago.  At the end of their paper, EPR note that while they take themselves to have shown that the standard quantum state ``does not provide a complete description of the physical reality,'' they have  ``left open the question of whether or not such a description exists.'' Nevertheless, they say, ``we believe \ldots\  that such a theory is possible.''

Costa de Beauregard himself saw his zigzag proposal as an \emph{objection} to the EPR argument. It showed how there might be spacelike influence, without action at a distance -- thus undermining EPR's main reason for assuming that a measurement choice at one location could not affect an ``element of reality'' at another location. In another sense, however, it amounts to a \emph{vindication} of EPR's conclusion. If accepted, it shows not only that EPR were right in thinking that the standard description is incomplete (because it leaves out the zigzag mechanism) but also that they were right in thinking that more complete  theory is indeed possible.

Our contribution here has been show how easy it is to motivate Costa de Beauregard's zigzag, via the symmetries underlying our use of mirror in Figure 2. We don't take ourselves to have offered conclusive arguments for the zigzag approach, of course, but we do urge that it deserves serious attention. For the moment, the prevailing view of entanglement -- that it involves the mysterious connections between real properties that Schr\"odinger derided as ``magic'' in 1935 -- seems to rest on a considerable failure of imagination. The Parisian zigzag offers an elegant alternative.\footnote{We gratefully acknowledge the assistance of comments from two referees. H.P.~also acknowledges the support of the TWCF research grant \emph{Information at the Quantum Physics/Statistical Mechanics Nexus.}} 



\makeatletter
\renewcommand\@biblabel[1]{#1. }
\makeatother



%


%

\end{document}